\documentclass{INTERSPEECH2023}
\usepackage{multicol}
\usepackage{multirow}
\usepackage{hyperref}
\usepackage{cleveref}
\usepackage{tabstackengine}
\usepackage{tabularx}
\usepackage{xcolor}
\usepackage{amsmath}
\usepackage{amssymb}% http://ctan.org/pkg/amssymb
\usepackage{pifont}% http://ctan.org/pkg/pifont
\usepackage{xurl}
\newcommand{\cmark}{\ding{51}}%
\newcommand{\xmark}{\ding{55}}%

\newcommand\Tstrut{\rule{0pt}{2.6ex}}         % = `top' strut
\newcommand\Bstrut{\rule[-0.9ex]{0pt}{0pt}}   % = `bottom' strut

\interspeechcameraready

\title{The ID R\&D VoxCeleb Speaker Recognition Challenge 2023\\System Description}

\name{Nikita Torgashov, Rostislav Makarov, Ivan Yakovlev,\\ Pavel Malov, Andrei Balykin, Anton Okhotnikov}
\address{ID R\&D Inc., New York, USA}
\email{\tt\{torgashov,makarov,yakovlev,pavel.malov,andrew.balykin,ohotnikov\}@idrnd.net}

\begin{document}

\maketitle

\begin{abstract}
This report describes ID R\&D team submissions for Track 2 (open) to the VoxCeleb Speaker Recognition Challenge 2023 (VoxSRC-23). 
Our solution is based on the fusion of deep ResNets and Self-supervised learning (SSL) based models trained on a mixture of a  VoxCeleb2 \cite{chung2018voxceleb2} dataset and a large version of a VoxTube \cite{yakovlev23_interspeech} dataset. The final submission to the Track 2 achieved the first place on the VoxSRC-23 public leaderboard with a $minDCF_{0.05}$ of 0.0762 and $EER$ of 1.30\%.

\end{abstract}
\vspace{0.3\baselineskip}

\section{Introduction}

In this paper, we present a detailed description of our solution for the VoxSRC-23, Track 2: fully supervised speaker verification, opened track. Our solution is built upon two popular families of neural network architectures for speaker recognition: ResNet and SSL-based models with WavLM, Unispeech, or XLSR feature extractors with ECAPA-TDNN model stacked on top of them. Given the unconstrained open track setup, we leveraged the usage of additional data for network training that yielded a significant performance boost. The final solution is an ensemble of the models scores together with the Quality Measurement Function (QMF) values fused using logistic regression. In the following sections, we provide a detailed description of our experiments and systems.

\section{Data}

\subsection{Train data}

To train the models we used the following training datasets:

\smallskip \noindent \textbf{VC2: VoxCeleb2} \cite{chung2018voxceleb2}. This is a base training dataset for most state-of-the-art speaker recognition models. It has great intra- and inter-speaker variability and arises from the same domain as challenge development and evaluation sets. 

\smallskip \noindent \textbf{VTL: VoxTube-Large}. We used a full version of the recently released and open-sourced VoxTube dataset \cite{yakovlev23_interspeech}. As VC2, this dataset is also collected from the video hosting platform YouTube. However, the collection process was based on the clustering of the pre-trained speaker embeddings without a face recognition model, and all the details could be found in \cite{yakovlev23_interspeech}.
While the open-sourced version of VoxTube contains more than 5K speakers, with almost 5K hours of speech in total, the VTL version is by a degree of magnitude bigger: it has more than 100K speakers with the same or greater number of sessions per speaker as in VoxTube and VC2 datasets. The publicly available version of VoxTube dataset can be found via link \footnote{\url{https://idrnd.github.io/VoxTube/}}.

\smallskip \noindent \textbf{VT30K: VoxTube-30K}. We have found out that the VTL dataset contains speakers that do not contribute significantly to model accuracy due to language or domain discrepancies towards the VC2 dataset. While such out-of-domain data can enhance generalization during the pre-training stage, it might hinder optimization during fine-tuning. To address this, we curated a subset of VTL speakers that are best aligned with the VoxCeleb domain. We derived the median speakers embeddings in both VoxCeleb1-dev and VTL datasets using a ResNet100 model pre-trained on VC2 data. With cosine similarity matching we identified the top-50 most similar VTL speakers for each speaker in VoxCeleb1-dev, ensuring no overlaps between datasets and removing any duplicates with a similarity above 0.8. This resulted in a refined subset of VTL termed as a "domain dataset filtering" (DDF).

\subsection{Validation data}
For validation, VoxCeleb1-test \cite{nagrani2017voxceleb} set and VoxSRC-20, 21, 22 \cite{nagrani2020voxsrc,brown2022voxsrc,huh2023voxsrc}, and 23 development sets were used.

\subsection{Augmentation data} \label{sec:data-aug}
For data augmentation during the initial training stage we used MUSAN \cite{snyder2015musan} and room impulse responses (RIR) \cite{ko2017study} databases. We used a standard augmentation strategy described in the training section of \cite{garcia2020magneto}. We also masked from 0 to 5 frames in the temporal axis and from 0 to 10 frames in the frequency axis using the SpecAug \cite{park2019specaugment}.

% For each training utterance, one of five various augmentation strategies was selected randomly:

% \begin{itemize}
%     \item {\bf Music}: A single music file is randomly selected from MUSAN and added to the original signal (5-15dB SNR). The duration of additive noise is matched to the duration of the original signal.
%     \item {\bf Noise}: Randomly selected noise from MUSAN added to the original recording (0-15dB SNR).
%     \item {\bf Speech}: Three to seven speakers are randomly picked, summed together, then added to the original signal (10-20dB SNR).
%     \item {\bf Reverb}: Artificially reverberate via convolution with real RIRs \cite{Szoke_2019}.
%     % \item {\bf Speed}: We applied a speed augmentation that increased a number of speakers in training data by a factor of 3. \cite{ko2015audio}.
%     \item {\bf SpecAugment}: We masked from 0 to 5 frames in the temporal axis and from 0 to 10 frames in the frequency axis using the SpecAug \cite{park2019specaugment}.
% \end{itemize}

% Architectures Table
\begin{table}[t]
    \centering
    \caption{ResNet-100 architecture}
    \label{tab:architectures}
    \scalebox{0.95}{
        \begin{tabular}{llr}
            \hline \hline
            \Tstrut\Bstrut
            \textbf{Layer name} &
            \multicolumn{1}{c}{\centering \textbf{Structure}} &
            \multicolumn{1}{p{2cm}}{\centering \textbf{Output} \\ (C $\times$ F $\times$ T)}
            \Tstrut\Bstrut \\
            \hline \hline
            \Tstrut\Bstrut
            Conv2D & \multicolumn{1}{c}{3$\times$3, 128, stride=1} & 128 $\times$ 96 $\times$ T \\
            \hline
            \rule{0pt}{20pt}ResBlock-1 &
            $\begin{bmatrix} 3 \times 3, 128 \\ 3 \times 3, 128 \\ \text{fwSE}, [128, 96] \end{bmatrix} \times 6$ &
            128 $\times$ 96 $\times$ T \\ [12pt]
            \hline
            \rule{0pt}{20pt}ResBlock-2 &
            $\begin{bmatrix} 3 \times 3, 128 \\ 3 \times 3, 128 \\ \text{fwSE}, [128, 48] \end{bmatrix} \times 16$ &
            128 $\times$ 48 $\times$ T/2 \\ [12pt]
            \hline
            \rule{0pt}{20pt}ResBlock-3 & 
            $\begin{bmatrix} 3 \times 3, 256 \\ 3 \times 3, 256 \\ \text{fwSE}, [128, 24] \end{bmatrix} \times 24$ &
            256 $\times$ 24 $\times$ T/4 \\ [12pt]
            \hline
            \rule{0pt}{20pt}ResBlock-4 & 
            $\begin{bmatrix} 3 \times 3, 256 \\ 3 \times 3, 256 \\ \text{fwSE}, [128, 12] \end{bmatrix} \times 3$ & 
            256 $\times$ 12 $\times$ T/8 \\ [12pt]
            \hline
            \Tstrut\Bstrut
            Flatten (C, F) & \multicolumn{1}{c}{---} & 3072 $\times$ T/8 \\
            CAS  & \multicolumn{1}{c}{---} & 6144 \\
            \hline
            \Tstrut\Bstrut
            Dense & \multicolumn{1}{c}{---} & 256 \\
            AM-Softmax & \multicolumn{1}{c}{---} & Num. of speakers \\
            \hline \hline
        \end{tabular}
    }
\end{table}

\begin{table*}[ht]
    \centering
    % \caption{ResNet100 models performance on the VoxSRC-23 dev set}
    \caption{Comparison of ResNet100 models performance on the VoxSRC-23 Dev set depending on the training and fine-tuning datasets. VC2 is VoxCeleb2, VTL - Large version of VoxTube dataset with more than 100k spks, and VT30K is a subset of VTL with 30k most relevant speakers.}
    \label{tab:resnet-comparison}
    \scalebox{0.9}{
    \begin{tabular}{ l c c c c }
        \hline
        \textbf{Model} &
        \textbf{Pretrain dataset} &
        \textbf{Finetune dataset} &
        \textbf{EER,\%} &
        \textbf{MinDCF}\\
        \hline
        % fwse_rn100m_vox2_spkin_l2e5_flr_v2_am03_nol2_v2
        RN1 & VC2       & VC2          & 3.24           & 0.174 \\
        
        % rn100_v016_am03_ptn_vox4_v2_flr_30k
        RN2 & VTL       & VT30K        & 2.73           & 0.156 \\
        
        % rn100fwse_v016_la_spd_dw_flr_noaug_am03_6s_v2-4sec_10crops
        RN3 & VTL + VC2 & VTL + VC2    & 2.54           & 0.141 \\
        
        % rn100_v016_am03_ptn_vox4_v2_flr_vox3-4sec_10crops
        RN4 & VTL       & VTL + VC2    & 2.18           & 0.123 \\
        
        % rn100_v016_am03_ptn_vox4_v2_flr_30k_1utt_v2-101ep-4sec_10crops
        RN5 & VTL       & VT30K + VC2  & \textbf{1.94}  & \textbf{0.105} \\
        
        \hline
    \end{tabular}}
\end{table*}

% Results Table
\begin{table*}[ht]
    \centering
    % \caption{Results on the VoxCeleb1-Cleaned and VoxSRC-20, 21, 22 and 23 dev sets}
    \caption{Evaluation results on the VoxCeleb1-Cleaned and VoxSRC-20, 21, 22 and 23 Dev sets. RNs are ResNet100 models trained on different subsets of data. The SSLs are ECAPA-TDNN models based on large SSL pre-trained backbones. SSL0 is the best open-source SSL model. SSL1-4 are our models trained on different subsets of data. All models are tested with a cosine similarity scoring without scores normalization or calibration.}
    \label{tab:results}
    \scalebox{.75}{
    \begin{tabular}{ l | c c | c c | c c | c c | c c | c c | c c }
        \hline \hline
        \Tstrut\Bstrut
        \multirow{2}{*}{\textbf{Model}} &
        \multicolumn{2}{c|} {\textbf{VoxCeleb1-O}} &
        \multicolumn{2}{c|} {\textbf{VoxCeleb1-E}} &
        \multicolumn{2}{c|} {\textbf{VoxCeleb1-H}} &
        \multicolumn{2}{c|} {\textbf{VoxSRC-20 Dev}} &
        \multicolumn{2}{c|} {\textbf{VoxSRC-21 Dev}} &
        \multicolumn{2}{c|} {\textbf{VoxSRC-22 Dev}} &
        \multicolumn{2}{c} {\textbf{VoxSRC-23 Dev}} \\
        &
        $EER[\%]$ & $DCF_{0.01}$ &
        $EER[\%]$ & $DCF_{0.01}$ &
        $EER[\%]$ & $DCF_{0.01}$ &
        $EER[\%]$ & $DCF_{0.05}$ &
        $EER[\%]$ & $DCF_{0.05}$ &
        $EER[\%]$ & $DCF_{0.05}$ &
        $EER[\%]$ & $DCF_{0.05}$
        \Tstrut\Bstrut \\
        \hline\hline
        \Tstrut\Bstrut
        
        RN1 & 0.43 & 0.043 & 0.65 & 0.070 &	1.24 & 0.122 & 2.14 & 0.114 & 2.62 & 0.149 & 1.67 &	0.103 & 3.24 & 0.174 \\

        RN2 & 0.23 & 0.033 & 0.52 & 0.047 & 1.01 & 0.092 & 1.76 & 0.083 & 2.02 & 0.136 & 1.53 &	0.087 & 2.73 & 0.156 \\

        RN3 & 0.23 & 0.020 & 0.49 & 0.047 &	0.93 & 0.088 & 1.67 & 0.081 & 1.89 & 0.117 & 1.35 &	0.077 & 2.54 & 0.141 \\

        RN4 & 0.19 & \textbf{0.010} & 0.40 & 0.038 & 0.81 & 0.074 & 1.46 & 0.073 & 1.74 & 0.110 & 1.11 &	0.066 & 2.18 & 0.123 \\

        \textbf{RN5} & \textbf{0.15} & 0.011 & \textbf{0.38} & \textbf{0.032} &	\textbf{0.74} & \textbf{0.064} & \textbf{1.31} & \textbf{0.064} & \textbf{1.43} & \textbf{0.088} & \textbf{1.04} &	\textbf{0.061} & \textbf{1.94} & \textbf{0.105} \\

        \hline

        SSL0 & 0.52 & 0.070 & 0.74 & 0.070 &  1.34 & 0.139 & 2.35 & 0.128 & 2.66 & 0.160 & 1.94 &	0.112 & 3.64 & 0.195 \\
        \textbf{SSL1} & \textbf{0.36} & \textbf{0.030} & \textbf{0.45} & \textbf{0.049} &	\textbf{0.93} & \textbf{0.089} & \textbf{1.72} & \textbf{0.089} & \textbf{1.86} & \textbf{0.108} & \textbf{1.31} &	\textbf{0.087} & \textbf{2.71} & \textbf{0.157} \\
        SSL2 & 0.38 & 0.042 & 0.59 & 0.063 & 1.19 & 0.116 & 2.14 & 0.111 & 2.45 & 0.146 & 1.61 &	0.106 & 3.45 & 0.182 \\
        SSL3 & 0.39 & 0.039 & 0.57 & 0.061 & 1.14 & 0.108 & 2.04 & 0.108 & 2.34 & 0.134 & 1.63 &	0.104 & 3.23 & 0.175 \\
        SSL4 & 0.41 & 0.049 & 0.54 & 0.060 & 1.06 & 0.109 & 1.89 & 0.103 & 2.09 & 0.132 & 1.48 &	0.099 & 3.00 & 0.164 \\
        % SSL4_{as} & 0.37 & 0.052 & 0.52 & 0.055 &	1.06 & 0.104 & 1.76 & 0.095 & 1.94 & 0.119 & 1.41 &	0.099 & 2.77 & 0.154 \\
        
        \hline \hline
        % \Tstrut\Bstrut
        % \textbf{Fusion} & 0.14 &	0.012&	0.36&	\textbf{0.035}&	\textbf{0.66}&	\textbf{0.060}&	\textbf{0.94}&	\textbf{0.056}  & 0.0 &	0.0 & 0.0 &	0.0 & 0.0 &	0.0\\
        
        % \hline \hline
    \end{tabular}}
\end{table*}

\section{Experiments}

\subsection{ResNets models}

We used a ResNet-34 \cite{garcia2020magneto} architecture as a baseline and applied a couple of modifications to the original architecture that led to ResNet with 100 hidden layers. As inputs for ResNet we used mean-normalized 96 Mel filter bank log-energies (MFB) with a 25 ms frame length, 10 ms step, and the FFT size of 512 over 20-7600 Hz frequency limits. Frequency-wise Squeeze-Excitation (fwSE) \cite{thienpondt2021integrating} blocks with bottleneck size 128 were added to the end of each residual module, and a Channel-dependent Attentive Statistics (CAS) \cite{desplanques2020ecapa} pooling was used. Details of the ResNet100 architecture are shown in table \ref{tab:architectures}.

\subsubsection{Initial training stage}
All models were trained using TensorFlow 2 framework \cite{abadi2016tensorflow} on Google Cloud TPUs. We trained all models for 300 epochs, 5000 steps each. The batch size was set to 256, and 4-second segments were randomly cropped for each utterance in the batch. We have also scheduled values of the learning rate, and a margin of the AM-Softmax \cite{wang2018additive} loss function, the scale parameter for which was set to 30. The learning rate scheduler had three phases: warmup, plateau, and decay. The learning rate was increased linearly from 1e-5 to 0.2, while the margin was equal to zero, for the first 10 epochs in the warmup phase. Then, the learning rate was fixed to 0.2 and the value of margin was linearly increased from 0 to 0.3 for the next 50 epochs in the plateau phase. After the margin achieved its maximum value, the learning rate decreased exponentially with a rate of 0.5 for every 20 epochs in the decay phase. For the data augmentation, we used strategies described in section \ref{sec:data-aug}. The L2 weights regularization was set to 1e-4.

We have tried two different data configurations for the model pre-training. The first one is a training on both, VTL and VC2 datasets, and the seconds one is a training on the VTL dataset only. While the first approach shows much higher accuracy, compared to the second one, the models pre-trained without the VC2 dataset show much better performance in the fine-tuning stage, where we trained on both datasets simultaneously. The results of these experiments (RN3 and RN4) can be found in the table \ref{tab:resnet-comparison}.

\subsubsection{Fine-tuning stage}

At the fine-tuning stage, we disabled all augmentations and decreased the value of L2-regularization to 1e-5. The number of epochs was decreased to 30 and 6-second training segments with a batch size of 160 were used. The learning rate was linearly increased from 1e-5 to 1e-2 for the first epoch, and then exponentially decreased with a rate of 0.5 each 5 epochs. The value of margin was fixed at value 0.3 for the whole stage.

\vspace{0.3\baselineskip}
We have used both, the VTL and the VT30K datasets for fine-tuning, and combined them with the VC2 dataset with equal sampling weights. The results of these experiments (RN4 and RN5) are presented in the table \ref{tab:resnet-comparison}. We have also tried to fine-tune the model on the VC2 dataset only, but found it less effective.

\subsection{SSL-based models}
% As second type of model architectures for speaker verification, SSL-based models like WavLM, 
As a second type of architectures used, we adopted SSL models for the speaker verification task. Self-supervised learning allows the model to learn useful representations from audio data without requiring explicit labels and recently SSL approaches have shown promising results for downstream tasks like speaker verification task. Due to the large size and high computing costs needed to train SSL models, the training was performed using spot TPU v2, and v3 accelerators provided by the Google Cloud platform. 

We followed the approach described in the WavLM paper using a stacked ECAPA-TDNN subnetwork on top of the SSL feature extractor. Our experiments involved multiple existing pre-trained and publicly available SSL backbones, including WavLM \cite{chen2022wavlm}, Unispeech \cite{chen2022unispeech}, and XLSR \cite{babu2021xls}.

\subsubsection{Common setup}
We trained the SSL models in 3 stages following the approach from the original WavLM paper \cite{chen2022wavlm}, ECAPA-TDNN with the number of channels C=1024 was used for all models. For all stages and models, we used SGD optimizer with momentum 0.9, AAM-Softmax loss \cite{deng2022aamsoftmax} with subcenters k=[1,3] and inter-top-k penalty \cite{zhao2021speakin}, and an exponential staircase with the warmup learning rate scheduler.  Here is the top-level overview of 3 training stages:

\smallskip \noindent \textbf{Stage 1: Pretraining ECAPA-TDNN}. In the first stage, we trained the ECAPA-TDNN weights only while keeping the SSL backbone frozen.

\smallskip \noindent \textbf{Stage 2: Fine-Tuning whole network}. Subsequently, we unfroze the SSL backbone and trained all the weights with a reduced learning rate.

\smallskip \noindent \textbf{Stage 3: Large Margin Fine-Tuning}. With all the weights unfrozen, we further refined the models by employing a large margin fine-tuning strategy, described in \cite{idlabvoxsrc20}. Margin was set to 0.5 and long 6 sec utterances were used for training, inter-top-k penalty was turned off at this stage.

% https://github.com/microsoft/unilm/issues/695#issuecomment-1110636164
\smallskip As a starting point for our experiments we adopted the hyperparameters detailed in a GitHub issue~\footnote{\url{https://github.com/microsoft/unilm/issues/695\#issuecomment-1110636164}} associated with the original publication. Our empirical observations suggest that SSL-based models are prone to significant overfitting when trained using small and medium-size datasets, like VoxCeleb2, and expanding training data multiple times gives better results and makes training more stable while tweaking hyperparameters, it also gives a possibility to reduce weight decay and to remove subcenters in AAM-Softmax loss.

\subsubsection{Models hyperparameters}

In this subsection in tables \ref{tab:hyperparameters_ssl_0} and \ref{tab:hyperparameters_ssl_1} we present detailed hyperparemters for best performing SSL models training. Some training stages setups differ from original training stages presented in WavLM paper \cite{chen2022wavlm}. Here is a list of models and their corresponding training hyperparameters:

\begin{itemize}
  \item \textbf{SSL0}: Open-source model by Microsoft. Best WavLM model from github repo \footnote{\url{https://github.com/microsoft/UniSpeech/tree/main/downstreams/speaker_verification}}
  \item \textbf{SSL1}: WavLM + ECAPA-TDNN, table \ref{tab:hyperparameters_ssl_0}
  \item \textbf{SSL2}: Unispeech + ECAPA-TDNN, table \ref{tab:hyperparameters_ssl_1}
  \item \textbf{SSL3}: XLSR + ECAPA-TDNN, table \ref{tab:hyperparameters_ssl_1}
  \item \textbf{SSL4}: WavLM + ECAPA-TDNN, table \ref{tab:hyperparameters_ssl_1}
\end{itemize}

\begin{table}[h]
    \centering
    \caption{Hyperparameters for 3 stages for SSL1 model. Learning rate scheduler setup is (gamma, number warmup epochs, number plateau epochs, number epochs per)}
    \scalebox{.7}{\begin{tabular}{l|ccc}
        % \hline
        \textbf{Parameter} & \textbf{Stage 1} & \textbf{Stage 2} & \textbf{Stage 3} \\
        \hline
        % \tabucline[1pt]
        SSL backbone &  & WavLM Large &  \\
        unfroze SSL & \xmark & \cmark & \xmark \\
        use augs & \cmark & \xmark & \xmark \\
        dataset & VC2:VT30K=1:1 & VC2:VT30K=1:1 & VC2:VT30K=1:1 \\
        max LR & 1.0 & 0.012 & 0.008 \\
        batch size & 2048 & 256 & 1280 \\
        utt len, sec & 3 & 3 & 6 \\
        max margin & 0.2 & 0.2 & 0.5 \\
        weight decay (L2) & 1e-4 & 2e-5 & 1e-5 \\
        Number of Epochs & 40 & 40 & 16 \\
        Steps per epoch & 1000 & 2000 & 1000 \\
        LR schedule  & (0.5, 2, 6, 2) & (0.6, 5, 3, 2) & (0.5, 2, 2, 2) \\
        \hline
        EER val vox1-test, \% & 0.71 & 0.54 & 0.48 \\
        % \hline
    \end{tabular}}
    \label{tab:hyperparameters_ssl_0}
\end{table}

\begin{table}[h]
    \centering
    \caption{Hyperparameters for 3 stages for SSL2-SSL4 models. Learning rate scheduler setup is (gamma, number warmup epochs, number plateau epochs, number epochs per)}
    \scalebox{.7}{\begin{tabular}{l|ccc}
        % \hline
        \textbf{Parameter} & \textbf{Stage 1} & \textbf{Stage 1} & \textbf{Stage 3} \\
        \hline
        % \tabucline[1pt]
        SSL backbone &  & Unispeech/XLSR/WavLM &  \\
        unfroze SSL & \xmark & \xmark & \cmark  \\
        use augs & \cmark & \xmark & \xmark \\
        dataset & VTL & VC2 & VC2 \\
        max LR & 1.0 & 0.45 & 0.008 \\
        batch size & 1024 & 1024 & 192 \\
        utt len, sec & 3 & 3 & 6 \\
        max margin & 0.2 & 0.2 & 0.5 \\
        weight decay (L2) & 1e-4 & 1e-6 & 1e-6 \\
        Number of Epochs & 30 & 40 & 20 \\
        Steps per epoch & 10000 & 2000 & 2000 \\
        LR schedule  & (0.5, 2, 10, 4) & (0.6, 2, 6, 2) & (0.5, 2, 2, 2) \\
        \hline
        EER val vox1-test, \% & 1.7 & 0.62 & 0.5 \\
        % \hline
    \end{tabular}}
    \label{tab:hyperparameters_ssl_1}
\end{table}

\section{Scoring and Fusion}

\subsection{Pairwise scoring and AS-Norm}

For inference, we sliced the input audios (both enrollment and verification) into $10 \times 4$ seconds chunks resulting in 100 cosine similarity scores in the same way as it was done in \cite{chung2018voxceleb2} and \cite{heo2020clova}. 
All the models results shown in the table \ref{tab:results} are given for a pairwise scoring technique.

Cosine similarity scores were further normalized by the application of an AS-Norm method. The AS-Norm cohort included all VoxCeleb2-dev speakers (mean embedding per speaker) with a preliminary subsampling of 20 utterances per speaker. To estimate mean and std of scores distribution for normalization \textit{top N = 100} trials were used. It is noteworthy that the normalization of the scores provided a good metric improvement on the VoxSRC-23 dev dataset and a much smaller improvement for the VoxCeleb1 dataset and VoxSRC development sets of previous years.

\subsection{Quality Measurement Functions}
For most of our submissions, we utilized the Quality Measuring Functions (QMFs) values as they usually give a huge performance boost, especially on VoxCeleb-based testing datasets \cite{zhao2021speakin},\cite{idlabvoxsrc20}. These are auxiliary measurements extracted over the input audio signal or utterance crops embeddings. They are assumed to provide additional information that is not captured by the single model utterance embedding. As a result, we exploited the following supplementary information for both enrollment and verification trials.

\vspace{0.5\baselineskip}
% \smallskip \noindent
\textbf{Audio quality measurements} included:
\begin{itemize}
    \item \textbf{NISQA} model \cite{mittag21_interspeech} speech quality perception values on the scale [1..5]: Mean Opinion Score (MOS), noisiness, discontinuity, coloration and loudness; 
    \item \textbf{Signal-to-Noise ratio} (SNR) estimation in dB  obtained from a Neural-based SNR estimator;
    \item \textbf{Babble Noise Detector} (BND) score indicating the probability of a background speech in audio obtained from a pre-trained neural network as well.
\end{itemize}

\vspace{0.5\baselineskip}
\textbf{Audio content attributes} included the estimates of:
\begin{itemize}
    \item \textbf{Age} and \textbf{gender} of a speaker;
    \item Neural VAD-based features: \textbf{speech length}, \textbf{file length};
    \item \textbf{Voice Liveness} probability from the SASV-like subnetwork system \cite{alenin2022subnetwork} representing a probability of an audio being replayed with any playback device.
\end{itemize}

\vspace{0.5\baselineskip}
\textbf{Model embedding} based features exploited the statistics of embeddings of crops used for a pairwise scoring, such as
\begin{itemize}
    \item \textbf{L1} and \textbf{L2 norms} of utterance mean embedding;
    \item \textbf{STD} of the mean embedding components across the dimensions axis;
    \item \textbf{MEAN} and \textbf{STD} values of STDs of each embedding dimension independently across the crops axis.
\end{itemize}

All extracted features were converted to either a binary or a real value format: e.g. feature engineering was applied to categorical features corresponding to gender and language match between the enrollment and verification trials that was encoded as a binary value based on the equality of estimated measurements. To some features, we applied a non-linear transformation for distribution normalization, e.g. logarithm to speech length or a file length feature. Finally, as a standardization technique, we applied a Min-Max normalization method per attribute to all real-valued features including models cosine scores and scores with AS-Norm.

\subsection{Evaluation metrics}
System's performance evaluation was conducted using the two metrics:
\begin{itemize}
    \item Minimum detection cost function \cite{nist2018} with parameters $P_{Target}=0.05$, $C_{Miss}=1$ and $C_{False Alarm}=1$,
    \item Equal Error Rate (EER) representing the operational point of equal False Acceptance (FA) and False Rejection (FR) error rates.
\end{itemize}

% \pagebreak

\subsection{Fusion scheme}
The output of our system is an implementation of a score-level linear fusion of normalized cosine similarity scores (with AS-Norm) for all the models and QMF values. To find the fusion weights and to map the output to [0..1] range we used the Logistic Regression with L1 penalty term from the sklearn \cite{sklearn_api} optimizing the error on VoxSRC-23 Dev set. The verification probability \(P\) and a logit score \(L\) were obtained according to the \cref{sigmoid} and \cref{fusion_score}:
\begin{equation} \label{sigmoid}
    \setstacktabbedgap{2pt}
    P(L) = \frac{1}{1 + e^{-L}},
\end{equation}
\begin{equation} \label{fusion_score}
    \setstacktabbedgap{2pt}
    L =  \bracketMatrixstack{w_1 & ... & w_n} \cdot 
            \bracketMatrixstack{S_1 \\ ... \\ S_n} + 
            \bracketMatrixstack{v_1 & ... & v_k} \cdot 
            \bracketMatrixstack{Q_1 \\ ... \\ Q_k},
\end{equation}
where \(w\) is a vector of models weights, \(S\) is a vector of normalized single models scores with AS-Norm, \(v\) is a vector of QMF weights and \(Q\) is a vector of QMF values.

\newpage

\section{Results analysis}

We can see from table \ref{tab:results} that our best SSL1 model outperforms open-source SotA model SSL0 by 20\% due to the train data used. Note that for a VoxSRC-23 challenge SSL-based models significantly lost their performance compared to ResNets (in the VoxSRC-22 challenge they were at the same level). It is clearly seen that trained in a supervised fashion ResNets provide 33\% metrics improvement compared to SSL-based systems when trained on the same datasets.

From the results in table \ref{tab:resnet-comparison} we can see that the model RN2 trained without the VoxCeleb2 dataset outperforms model RN1, trained on the VoxCeleb2 dataset only, by 10\% relative. We can also see that the addition of the VoxCeleb2 dataset to the fine-tuning stage allowed us to get up to 20\% relative improvement and to achieve the best results.
This table also shows that changing the pre-training strategy from joint training on VoxCeleb2 and VoxTube-Large (RN3) to only VoxTube-Large training, results (RN4) in a 12\% relative performance boost, considering the subsequent fine-tuning.
Also, the reduction of the domain mismatch between VoxTube-Large and VoxCeleb1 by a DDF technique improved the overall performance of the models.

And lastly, we have found that usage of QMFs can tremendously improve the system quality (see table \ref{tab:final-results}). In particular, we see a huge boost from using the model embedding based QMFs, as these values were extracted over the 10 crops of one utterance, and they capture the dynamics of embedding over the time. Moreover, we have found L1 regularization crucial to enhance our fusion performance, considering its property to implicitly conduct feature selection. Our final fusion consists of 10 single models presented in the table \ref{tab:results}. Table \ref{tab:final-results} shows the results on VoxSRC-23 dev and eval sets for our best single model RN5 with cosine pairwise scores only, and our fusion of 10 models with AS-Norm and various QMF values applied.

\begin{table}
    \centering
    % \caption{Results on the VoxSRC-23 dev and eval sets}
    \caption{Evaluation results of four submissions on VoxSRC-23 Dev and VoxSRC-23 Eval sets: the best single ResNet100 model with cosine scoring (a), fusion of all models with AS-Norm (b), fusion b and the embedding based QMF (c), and the best fusion with all QMFs (d).}
    \label{tab:final-results}
    \scalebox{0.8}{
    \begin{tabular}{ l | c c | c c }
        \hline \hline
        \Tstrut\Bstrut
        \multirow{2}{*}{\textbf{Model}} &
        \multicolumn{2}{c|} {\textbf{VoxSRC-23 Dev}} &
        \multicolumn{2}{c} {\textbf{VoxSRC-23 Eval}} \\
        &
        $EER[\%]$ & $DCF_{0.05}$ &
        $EER[\%]$ & $DCF_{0.05}$
        \Tstrut\Bstrut \\
        \hline\hline
        \Tstrut\Bstrut
        
        \textit{a.} RN5               & 1.94          & 0.105          & 2.14          & 0.110 \\

        \textit{b.} Fusion            & 1.45          & 0.086          & 1.88          & 0.096 \\
        
        \textit{c.} Fusion + emb. QMF & 1.06          & 0.069          & 1.38          & 0.078 \\

        \textit{d.} Fusion + all QMF  & \textbf{0.94} & \textbf{0.056} & \textbf{1.30} & \textbf{0.076} \\
        
        \hline \hline
    \end{tabular}}
\end{table}

\section{Conclusions}
In this report, we presented our solution for Track 2 of the VoxSRC-23 challenge. We have found the significant importance of the DDF technique and the usage of QMF values in fusion. We also observed a positive trend in extending the amount of training speech data for the open Track 2, as our ResNet100, trained on a mixture of VoxCeleb2-dev and VoxTube-Large, achieves state-of-the-art performance on the VoxCeleb1-test protocols. In future work, we would like to reach the supervised model quality with our SSL-based models. We would also like to pre-train SSL models using a mixture of VoxCeleb2-dev and VoxTube-Large datasets.

\clearpage

\bibliographystyle{IEEEtran}
\bibliography{voxsrc23}

\end{document}